# STOCHASTIC DYNAMICS OF STORM SURGE WITH STABLE NOISE

**Joshua Frankie Rayo\* and Vena Pearl Boñgolan**

Department of Computer Science, College of Engineering

University of the Philippines Diliman, Quezon City, Philippines

e-mail: *jbrayo@up.edu.ph, bongolan@up.edu.ph

## Abstract

The Advanced Circulation (ADCIRC) and Simulating Nearshore Waves (SWAN) coupled model is modified to include a stochastic term in the shallow water equations that represents random external forces from debris carried by surge and short-term local scale atmospheric fluctuations. We added $\alpha$-stable noise, uncorrelated in space and time, in the forcing terms of the coupled model. Inputs to the model are unstructured computational mesh derived from topography and bathymetry, land cover classification, tidal potential constituents and atmospheric forcing. The model simulated surge height of around five meters rushing at four meters per second near Tacloban City downtown. Underestimation of simulated surge height is expected with use of bare earth model and absence of fluid sources on the governing equations, while overestimation of simulated peak height also occurs due to presence of concrete barriers that reduced surge height and inundation extent. The stochastic model is sensitive to random external forces during low tide and relatively higher fluid speed. Low tide happens when the fluid speed is maximum and water elevation is lowest, while higher fluid speed is brought by external forces like storm surge. However, the difference of stochastic solutions from the deterministic solution averages to zero and there is no significant improvement of the storm surge model in general when it comes to additive noise. This is an expected result since the noise used has zero mean. As $\alpha$ goes to zero larger jumps occur more frequently so $\sigma$ needs to be as small as $10^{-8}$ for simulation stability.

# Introduction

A storm surge model may consider random external forces for realistic simulations. This can come from debris carried by surge, short-term local scale atmospheric fluctuations, electromagnetic hydrodynamic forces and others. However, until today there is no computational fluid dynamics (CFD) software that can do it. This research is inspired by recent results regarding solutions of stochastic Navier-Stokes equations (SNS).

Hanggi and Jung describes external noise as "imposed on some subsystem by a larger fluctuating environment in which the subsystem is immersed" [1]. The subsystem can resemble fluid circulation, while the larger environment can be winds, tidal potential forces, earth's movement & electromagnetic activity. External noise can resemble the reaction of fluid to soil and structural debris which is not considered in the subsystem. The Navier-Stokes (NS) equations explain the velocity profile of a fluid flowing in a domain subject to an external body force:

$$\frac{\partial \vec{u}}{\partial t} + \vec{u} \cdot \nabla \vec{u} + \nabla p - v \Delta \vec{u} = \vec{F}(x,t) + \vec{\xi}(t) \qquad \operatorname{div} \vec{u} = \mathbf{0}$$

where $\vec{u}$ is fluid velocity, $p$ is fluid pressure, $v$ is kinematic viscosity, $\xi$ is additive noise, $x$ is space and $t$ is time. Flandoli [2] recommends the use of additive noise when there is a Laplacian term $v\Delta\vec{u}$. Additive noise is used in turbulence investigation when energy balance is contained, and average energy introduced to the system is known beforehand [2]. Menaldi and Sritharan [5] cited that NS equation is a widely used model for atmospheric and ocean dynamics. Their paper provides a proof for the existence and uniqueness of strong solutions for 2D SNS with Gaussian space-time white noise in bounded and unbounded domains. Addition of noise reduces the number of physically meaningful invariant measures or solutions since it is not unique for the deterministic case [5]. Brzeźniak, Goldys and Le Gia [3] analyzed the movement of fluids on a spherical shell using Navier-Stokes equations on a 2-dimensional rotating sphere with space-correlated but time uncorrelated Gaussian noise. Existence and uniqueness of pathwise variational solutions is proved and an asymptotically compact random dynamical system was shown which is crucial for the existence of a compact random attractor and an invariant measure [3]. Well-posedness [4] as well as existence and uniqueness [6] of the solution for 2D SNS with $\alpha$-stable process was also obtained with unique invariant measure.

Inspired by the generalized Central Limit Theorem, $\alpha$-stable process is used in modeling random variables. It can accommodate heavy tails and skewness to cover extreme events that can have a good fit to observed data [7]. It is denoted by $S_X(\alpha, \beta, \gamma, \delta)$ where $\alpha \in [0, 2.0)$ is index of stability, $\beta \in [-1.0, 1.0]$ is skewness parameter, $\sigma > 0$ is variance (scale factor) and $\delta \in \mathbb{R}$ is location parameter (mean value when $\alpha \in [1.0, 2.0)$). The fastest and most accurate algorithm for simulating $\alpha$-stable processes was given by Chambers, Mallows and Stuck (CMS). This was improved afterwards by McCulloch [8].

Nonlinear and statistical storm surge models are used without making the governing equations stochastic. Nonlinear predictive chaotic models are built from time series data to predict storm surge dynamics [9]. Reconstruction of large number of synthetic, physically consistent storm surge events are generated from sea level observation from tide gauges [10]. Hypothetical storms are generated by permuting properties of hurricanes to calculate exceedance probability of storm surge with certain height in a location [11].

ADCIRC is a "highly-developed system of computer programs for solving time dependent, free surface circulation and transport problems in two depth-integrated or three dimensions" [12]. Applications of ADCIRC include storm surge and flooding simulation, tide and wind driven circulation modeling, scalar transport, among others. Its governing equations are formulated using hydrostatic pressure and Boussinesq approximations on shallow water equations. Galerkin finite-element method is used to compute fluid velocity     by solving the 2D depth-integrated momentum equations:

$$\frac{\partial U}{\partial t} + U\frac{\partial U}{\partial x} + V\frac{\partial U}{\partial y} - fV = -g\frac{\partial\left[\zeta+\frac{P_S}{g\rho_0}-\alpha\eta\right]}{\partial x} + \frac{\tau_{sx}}{H\rho_0} - \frac{\tau_{bx}}{H\rho_0} + \frac{M_x}{H} - \frac{D_x}{H} - \frac{B_x}{H} \quad (2.1)$$

$$\frac{\partial V}{\partial t} + U\frac{\partial V}{\partial x} + V\frac{\partial V}{\partial y} + fU = -g\frac{\partial\left[\zeta+\frac{P_S}{g\rho_0}-\alpha\eta\right]}{\partial y} + \frac{\tau_{sy}}{H\rho_0} - \frac{\tau_{by}}{H\rho_0} + \frac{M_y}{H} - \frac{D_y}{H} - \frac{B_y}{H} \quad (2.2)$$

Surface elevation is obtained by solving the generalized wave continuity equation:

$$\frac{\partial H}{\partial t} + \frac{\partial}{\partial x}(UH) + \frac{\partial}{\partial y}(VH) = 0 \quad (3)$$

using Galerkin finite-element method as well, keeping all nonlinear terms retained. SWAN is a "third-generation wave model that computes short-crested, wind-generated waves in water bodies". It treats short waves as energy spectrum and apply conservation of wave action density to model wave-current interactions. Its numerical solution was implemented using Gauss-Seidel sweeping technique [12]. Tight coupling of ADCIRC and SWAN models increases efficiency and is highly scalable in a parallel computing infrastructure. This computational paradigm is helpful for early warning efforts, where time is essential [13].

We modified an existing model to incorporate random external forces and account for measurement errors on input data. $\alpha$-stable process, space-time uncorrelated is added as noise on the numerical solution of ADCIRC and SWAN coupled model. Model performance is validated with storm surge happened in Tacloban City during the landfall of typhoon Haiyan on November 2013 based on field surveys.

**Methodology**

**Incorporating Additive Stable Noise**

Noise term was added on the right side of equations 2.1, 2.2 to represent random external force. The source code was modified as follows:

1. For each timestep

a. Generate two random stable numbers $Y_1, Y_2 \sim S_X(\alpha, \beta = 0, \sigma, \delta = 0)$ using McCulloch's improvement of CMS algorithm [8]. This resembles additive noise for horizontal coordinates $x$ and $y$.

b. Add them to force vector before calculating velocity of a computational node.

2. Repeat steps above until simulation ends.

McCulloch's algorithm is described below. With valid inputs on $\alpha, \beta, \sigma, \delta$ [8],

1. Generate random variable $V$ uniformly distributed on $\left(-\frac{\pi}{2}, \frac{\pi}{2}\right)$ and independent exponential random variable $W$ with mean 1.

2. If $\alpha = 2.0$ then $X = 2\sqrt{W} \sin V$.

3. Else if $\alpha = 1.0$ and $\beta = 0$ then $X = \tan V$.

4. Else if $\alpha = 1.0$ and $\beta \neq 0$ then $X = \frac{2}{\pi}\left\{\left(\beta V + \frac{\pi}{2}\right)\tan V - \beta \log\left[\frac{\frac{\pi}{2}W \cos V}{2\left(\beta V + \frac{\pi}{2}\right)}\right]\right\}$.

5. Else if $\alpha \neq 1.0$ then $X = S_{\alpha,\beta} \cdot \left\{\frac{\sin[\alpha(V + B_{\alpha,\beta})]}{(\cos V)^{1/\alpha}}\right\} \cdot \left\{\frac{\cos[V - \alpha(V + B_{\alpha,\beta})]}{W}\right\}^{1/\alpha - 1}$, where $S_{\alpha,\beta} = \left(1 + \beta^2 \tan^2 \frac{\pi\alpha}{2}\right)^{\frac{1}{2\alpha}}$ and $B_{\alpha,\beta} = \frac{\arctan\left(\beta \tan\frac{\pi\alpha}{2}\right)}{\alpha}$

6. Since $X \sim S_X(\alpha, \beta, \sigma = 1.0, \delta = 0)$ a linear transformation from $X$ to $Y \sim S_X(\alpha, \beta, \sigma, \delta)$ is computed as $Y = \begin{cases} \sigma X + \delta, & \alpha \neq 1 \\ \sigma X + \frac{2}{\pi}\beta\sigma \log \sigma + \delta, & \alpha = 1 \end{cases}$

7. Return $Y$

Generating a random stable number took 4 milliseconds on an Intel i7 2.80GHz processor. To investigate the effects of noise with respect to the deterministic model, different parameter values were assigned on random number generation: $\alpha = 0.75, 1.00, 1.25, 1.50, 1.75, 2.00$ and $\sigma = 10^{-10}, 10^{-9}, \dots, 10^{-3}$. For each scenario 30 simulations were run and number of successful simulations (did not go unstable) were recorded. Time-series difference from the deterministic model were also computed.

**Unstructured Grid**

Topography and bathymetry were obtained from Phil-LiDAR [14] as digital terrain model (DTM) with spatial resolution of one meter, while coastline was obtained from OpenStreetMaps. Quantum GIS software was used to visualize acquired data, project them to WGS84 coordinate system and extract computational domain boundaries. DTM is a bare earth model that does not include man-made structures. This raster format was converted to XYZ text format that contains x-y geographic coordinates and height/depth of terrain.

The coastline shapefile was simplified to reduce tiny complex details and converted afterwards to XYZ format of WGS84 projection. This is the set of points to be forced as nodes in the finite element mesh. Linear truncation error analysis toolbox of Surface Water Modeling System was used to compute optimal element sizes and distribute truncation error uniformly across the computational mesh. This toolbox is of great help in generating the mesh quickly with minimum user interaction [14]. The Courant-Friedrichs-Lewy (CFL) number, a numerical stability criterion, is kept as low as possible to accommodate effects of noise. This implies that space discretization $\Delta x$ should be long enough to cover the distance traveled by a fluid parcel in one timestep $\Delta t$.

**Boundary Forcing**

Thirteen major tidal harmonic constituents (table 1) were used to force elevation in the open boundaries:

**Table 1. Thirteen major tidal constituents that can be used for tidal prediction.**

| Darwin Symbol | Species |
|---|---|
| $K_1$ | Lunar diurnal |
| $O_1$ | Lunar diurnal |
| $P_1$ | Solar diurnal |
| $Q_1$ | Larger lunar elliptic diurnal |
| $N_2$ | Larger lunar elliptic semidiurnal |
| $M_2$ | Principal lunar semidiurnal |
| $S_2$ | Principal solar semidiurnal |
| $K_2$ | Lunisolar semidiurnal |
| $M_4$ | Shallow water overtides of principal lunar |
| $MS_4$ | Shallow water quarter diurnal |
| $MN_4$ | Shallow water quarter diurnal |
| $MM$ | Lunar monthly |
| $MF$ | Lunisolar fortnightly |

Amplitudes and phases were obtained from TOPEX/Poseidon Global Inverse Solution version 8 model [15] with grid resolution of 1/30 degrees. In land boundaries external overflow flux condition was used; if water goes beyond the boundary height, it will flow out of the domain.

**Nodal Attributes**

Nationwide land cover data was published by National Mapping and Resource Information Agency (NAMRIA) with 14 categories (see figure 1). Since associated surface roughness coefficients were not present in their resource, their closest counterparts from NLCD (National Land Cover Database) were used [16]. Directional surface roughness length specified by Mattocks and Forbes that reduces wind speed based on land cover [16] was also introduced. Horizontal eddy viscosity was set to 5 $m/s^2$ and 50 $m/s^2$ in water and land, respectively [16]. Spatially varying bottom friction for computational nodes underwater was computed using the hybrid quasi-linear formulation specified by Dietrich, et. al [12].

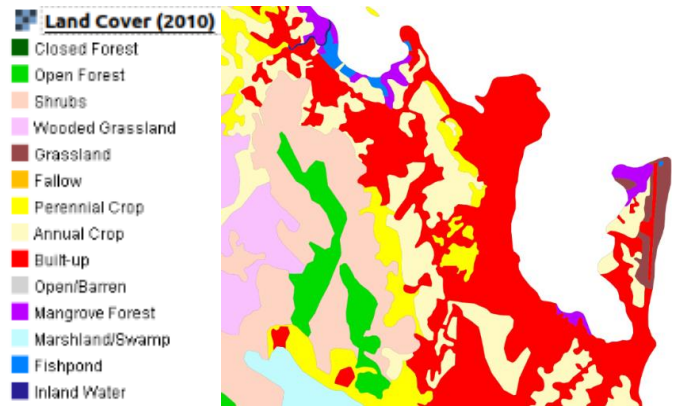

Figure 1. Land cover data of Tacloban City (11.25° N, 125° E)

**Model Control**

Timestep was set to no more than one second for simulation stability described by CFL condition. Output data including fluid velocity, elevation and wind speed were recorded every 300 timesteps to reduce disk space consumption and execution time. To implement the wetting and drying algorithm, minimum cutoff depth was set to $H_0 = 0.10\ m$ and minimum velocity was set to $U_{min} = 0.01\ m/s$ [12]. Characteristics of major tidal potential constituents such as amplitude, period, nodal coefficient and equilibrium argument were also included in the model. Earth tide reduction / elasticity factors was obtained from Wahr [17].

**Atmospheric Forcing**

The Best Track Format published by Joint Typhoon Warning Center (JTWC) [18] was used for atmospheric forcing in the storm surge model (see figures 2 and 3). Asymmetric Holland model is used to interpolate the wind velocity and atmospheric pressure, while Garret's formula is used to compute wind stress from wind velocity [16].

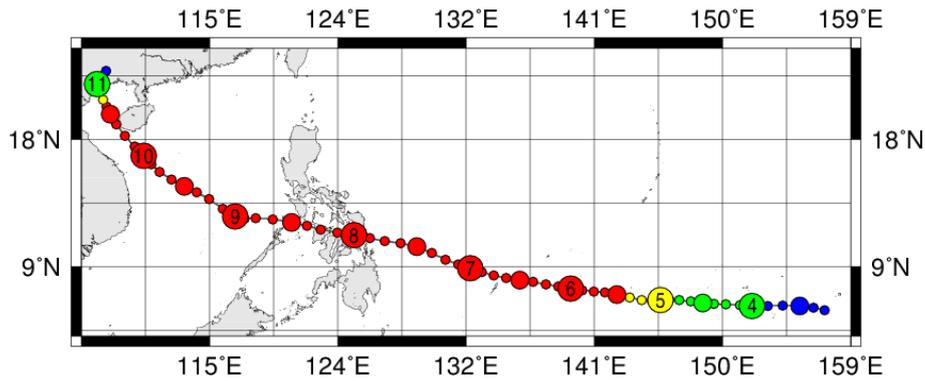

Figure 2. Typhoon Haiyan track visualization from Japan National Institute of Informatics (http://agora.ex.nii.ac.jp/digital-typhoon/summary/wnp/l/201330.html.en). Numbers in circles indicate the specific day of November 2013. Colors indicate categorical strength of the typhoon, with red being Category 5.

```
WP, 31, 2013110606,  , BEST,   0,  76N, 1380E, 135, 922, ST, 64, NEQ,  35,  35,  35,  35, 1006, 220,  7,  0,  5, W, 0, , 0, 0,  HAIYAN, D,
WP, 31, 2013110612,  , BEST,   0,  79N, 1361E, 150, 911, ST, 34, NEQ, 125, 120, 120, 125, 1006, 200,  7,  0,  5, W, 0, , 0, 0,  HAIYAN, D,
WP, 31, 2013110612,  , BEST,   0,  79N, 1361E, 150, 911, ST, 50, NEQ,  75,  75,  75,  75, 1006, 200,  7,  0,  5, W, 0, , 0, 0,  HAIYAN, D,
WP, 31, 2013110612,  , BEST,   0,  79N, 1361E, 150, 911, ST, 64, NEQ,  45,  45,  45,  45, 1006, 200,  7,  0,  5, W, 0, , 0, 0,  HAIYAN, D,
WP, 31, 2013110618,  , BEST,   0,  82N, 1343E, 155, 907, ST, 34, NEQ, 125,  95, 110, 125, 1006, 200, 12,  0, 15, W, 0, , 0, 0,  HAIYAN, D,
WP, 31, 2013110618,  , BEST,   0,  82N, 1343E, 155, 907, ST, 50, NEQ,  60,  55,  50,  65, 1006, 200, 12,  0, 15, W, 0, , 0, 0,  HAIYAN, D,
WP, 31, 2013110618,  , BEST,   0,  82N, 1343E, 155, 907, ST, 64, NEQ,  45,  30,  25,  45, 1006, 200, 12,  0, 15, W, 0, , 0, 0,  HAIYAN, D,
WP, 31, 2013110700,  , BEST,   0,  87N, 1327E, 155, 907, ST, 34, NEQ, 125,  95, 110, 125, 1006, 200, 15,  0, 20, W, 0, , 0, 0,  HAIYAN, D,
WP, 31, 2013110700,  , BEST,   0,  87N, 1327E, 155, 907, ST, 50, NEQ,  60,  55,  50,  65, 1006, 200, 15,  0, 20, W, 0, , 0, 0,  HAIYAN, D,
WP, 31, 2013110700,  , BEST,   0,  87N, 1327E, 155, 907, ST, 64, NEQ,  45,  30,  25,  45, 1006, 200, 15,  0, 20, W, 0, , 0, 0,  HAIYAN, D,
WP, 31, 2013110706,  , BEST,   0,  94N, 1310E, 160, 903, ST, 34, NEQ, 125,  95, 110, 125, 1003, 200, 17,  0, 25, W, 0, , 0, 0,  HAIYAN, D,
WP, 31, 2013110706,  , BEST,   0,  94N, 1310E, 160, 903, ST, 50, NEQ,  60,  55,  50,  65, 1003, 200, 17,  0, 25, W, 0, , 0, 0,  HAIYAN, D,
WP, 31, 2013110706,  , BEST,   0,  94N, 1310E, 160, 903, ST, 64, NEQ,  45,  30,  25,  45, 1003, 200, 17,  0, 25, W, 0, , 0, 0,  HAIYAN, D,
WP, 31, 2013110712,  , BEST,   0, 102N, 1290E, 170, 895, ST, 34, NEQ, 130, 100, 115, 130, 1003, 200, 17,  0, 25, W, 0, , 0, 0,  HAIYAN, D,
WP, 31, 2013110712,  , BEST,   0, 102N, 1290E, 170, 895, ST, 50, NEQ,  65,  60,  55,  70, 1003, 200, 17,  0, 25, W, 0, , 0, 0,  HAIYAN, D,
WP, 31, 2013110712,  , BEST,   0, 102N, 1290E, 170, 895, ST, 64, NEQ,  50,  35,  30,  50, 1003, 200, 17,  0, 25, W, 0, , 0, 0,  HAIYAN, D,
WP, 31, 2013110718,  , BEST,   0, 106N, 1269E, 170, 895, ST, 34, NEQ, 130, 115, 120, 130, 1000, 220, 17,  0, 25, W, 0, , 0, 0,  HAIYAN, D,
WP, 31, 2013110718,  , BEST,   0, 106N, 1269E, 170, 895, ST, 50, NEQ,  65,  60,  60,  70, 1000, 220, 17,  0, 25, W, 0, , 0, 0,  HAIYAN, D,
WP, 31, 2013110718,  , BEST,   0, 106N, 1269E, 170, 895, ST, 64, NEQ,  50,  45,  40,  50, 1000, 220, 17,  0, 25, W, 0, , 0, 0,  HAIYAN, D,
WP, 31, 2013110800,  , BEST,   0, 110N, 1247E, 165, 899, ST, 34, NEQ, 130, 115, 120, 130, 1000, 220, 15,  0, 20, W, 0, , 0, 0,  HAIYAN, D,
WP, 31, 2013110800,  , BEST,   0, 110N, 1247E, 165, 899, ST, 50, NEQ,  65,  60,  60,  70, 1000, 220, 15,  0, 20, W, 0, , 0, 0,  HAIYAN, D,
WP, 31, 2013110800,  , BEST,   0, 110N, 1247E, 165, 899, ST, 64, NEQ,  50,  45,  40,  50, 1000, 220, 15,  0, 20, W, 0, , 0, 0,  HAIYAN, D,
WP, 31, 2013110806,  , BEST,   0, 114N, 1225E, 145, 914, ST, 34, NEQ, 130, 115, 120, 130, 1002, 200, 10,  0, 10, W, 0, , 0, 0,  HAIYAN, D,
WP, 31, 2013110806,  , BEST,   0, 114N, 1225E, 145, 914, ST, 50, NEQ,  65,  60,  60,  70, 1002, 200, 10,  0, 10, W, 0, , 0, 0,  HAIYAN, D,
WP, 31, 2013110806,  , BEST,   0, 114N, 1225E, 145, 914, ST, 64, NEQ,  50,  45,  40,  50, 1002, 200, 10,  0, 10, W, 0, , 0, 0,  HAIYAN, D,
WP, 31, 2013110812,  , BEST,   0, 118N, 1204E, 130, 926, ST, 34, NEQ, 130, 115, 120, 130, 1004, 200, 15,  0, 20, W, 0, , 0, 0,  HAIYAN, D,
WP, 31, 2013110812,  , BEST,   0, 118N, 1204E, 130, 926, ST, 50, NEQ,  65,  60,  60,  70, 1004, 200, 15,  0, 20, W, 0, , 0, 0,  HAIYAN, D,
WP, 31, 2013110812,  , BEST,   0, 118N, 1204E, 130, 926, ST, 64, NEQ,  50,  45,  40,  50, 1004, 200, 15,  0, 20, W, 0, , 0, 0,  HAIYAN, D,
WP, 31, 2013110818,  , BEST,   0, 124N, 1179E, 120, 933, TY, 34, NEQ, 130, 120, 120, 130, 1004, 200, 10,  0,  0, W, 0, , 0, 0,  HAIYAN, D,
WP, 31, 2013110818,  , BEST,   0, 124N, 1179E, 120, 933, TY, 50, NEQ,  65,  65,  65,  65, 1004, 200, 10,  0,  0, W, 0, , 0, 0,  HAIYAN, D,
```

Figure 3. Typhoon Haiyan best track data, from Japan National Institute of Informatics (http://agora.ex.nii.ac.jp/digital-typhoon/summary/wnp/l/201330.html.en).

## Results

**Tacloban City Storm Surge**

*Characteristics of Storm Surge*

Simulation was run from November 04, 2013 02:00 UTC+8 to November 08, 2013 02:00 UTC+8 with timestep of one second. Total area covered was about 76 million square meters that generated a mesh with 2985 nodes and 5776 elements. Mesh resolution varies from 31 meters up to 1434 meters.

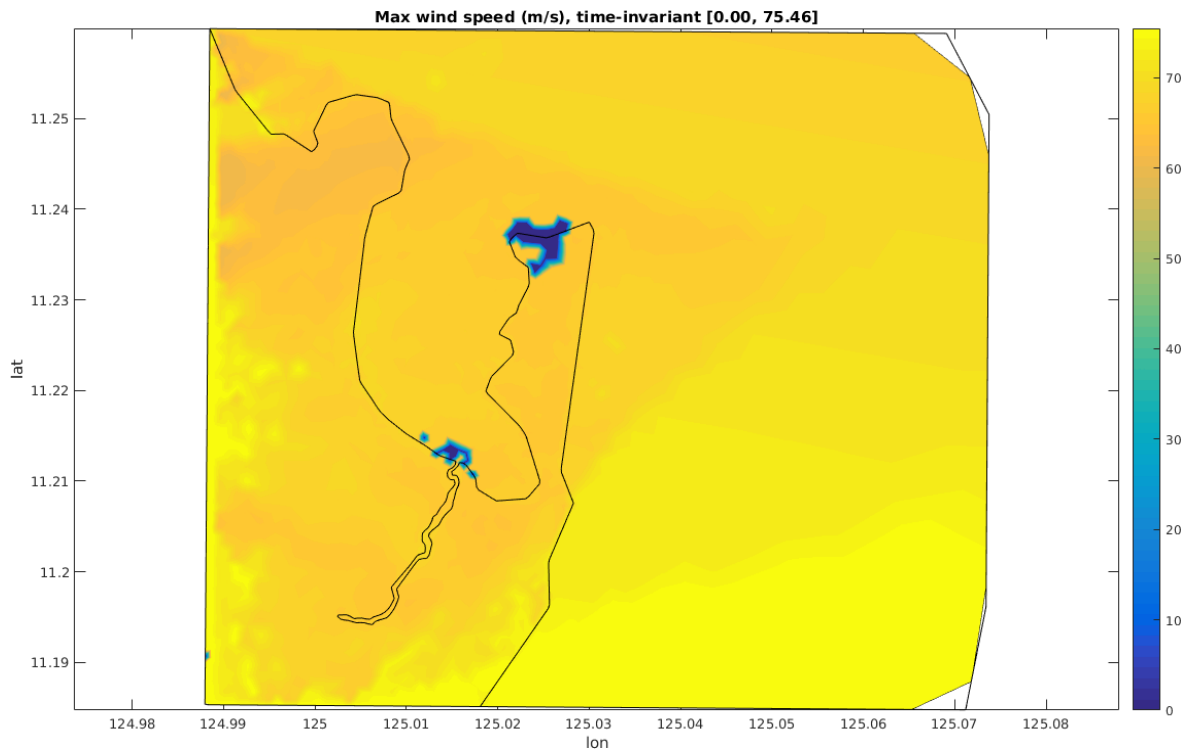

Figure 4. Simulated max wind speed on Earth's surface, time invariant during the onset of typhoon Haiyan.

Figure 4 shows maximum wind speed on Earth's surface during the typhoon. It reached 75 kilometers per hour on the ocean (right side) and gradually reduced to around 60 kilometers per hour on land (left side). Upwind wind speed was modulated by including land cover attributes. Mangrove forests (see figure 1) that protects water from wind stress have zero wind speed. At November 8, 2013 06:50 UTC+8 strongest winds (eyewall) are already present at Tacloban City with speed of 75 kilometers per hour. At 7:05 UTC+8 atmospheric pressure was around 965 mb. At 07:25 UTC+8 the eyewall moves closer toward Tacloban City. At 08:45 UTC+8 there were no typhoon winds present. Ground observations of Morgerman [20] on atmospheric conditions were captured by the model in good agreement.

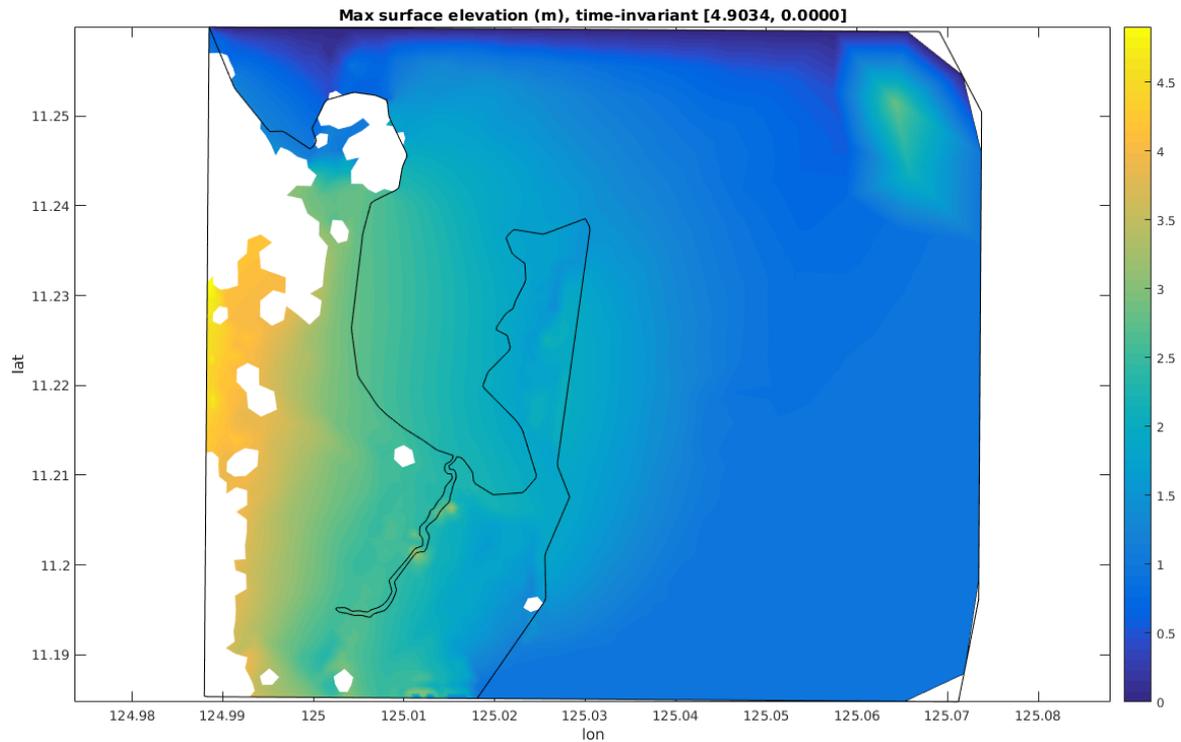

Figure 5. Simulated maximum surge height and inundation extent.

Simulated surge height of around five meters underestimated eyewitness reports surveyed by Soria, et. al. [22]. Furthermore, observed heights and inundation extent was larger than what was surveyed by Tajima, et. al. [20]. This is due to bare land representation of topography used that allowed water to flow further inland. Structures and debris can make water molecules accumulate on top of each other thus creating higher surge and may cover smaller inundation extent, due to conservation of mass. Precipitation is also not included in the model. Flow speed of about four meters per second was experienced near shore.

*Sensitivity Analysis to Additive Noise*

Figures 6 and 7 shows time-series difference from deterministic solution on water elevation and fluid speed when $\alpha = 2.00$, $\sigma = 10^{-3}$ (Gaussian). It was observed that the magnitude of difference in elevation becomes larger during low tide and higher fluid speed.

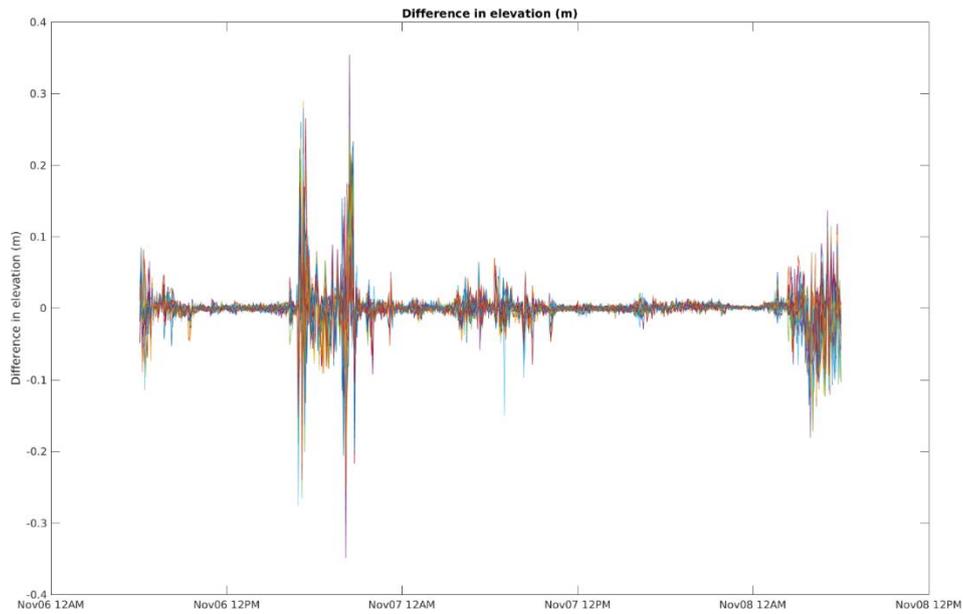

Figure 6. Stochastic simulations with additive noise $\alpha = 2.00$, $\gamma = 10^{-3}$ and its respective differences on time-series water elevation from deterministic model. Each line represents a successful simulation.

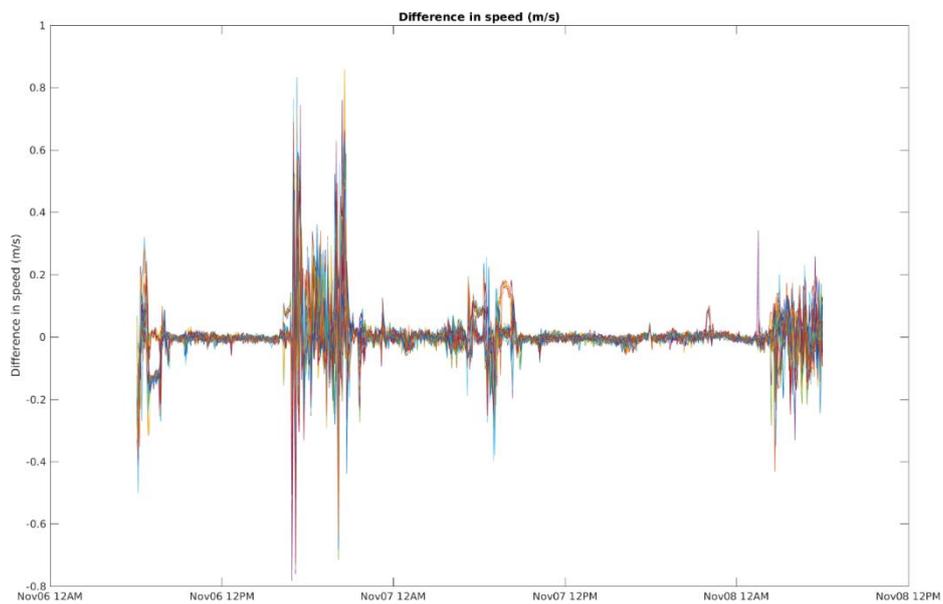

Figure 7. Stochastic simulations with additive noise $\alpha = 2.00$, $\gamma = 10^{-3}$ and its respective differences in time-series fluid speed from deterministic model. Each line represents a successful simulation.

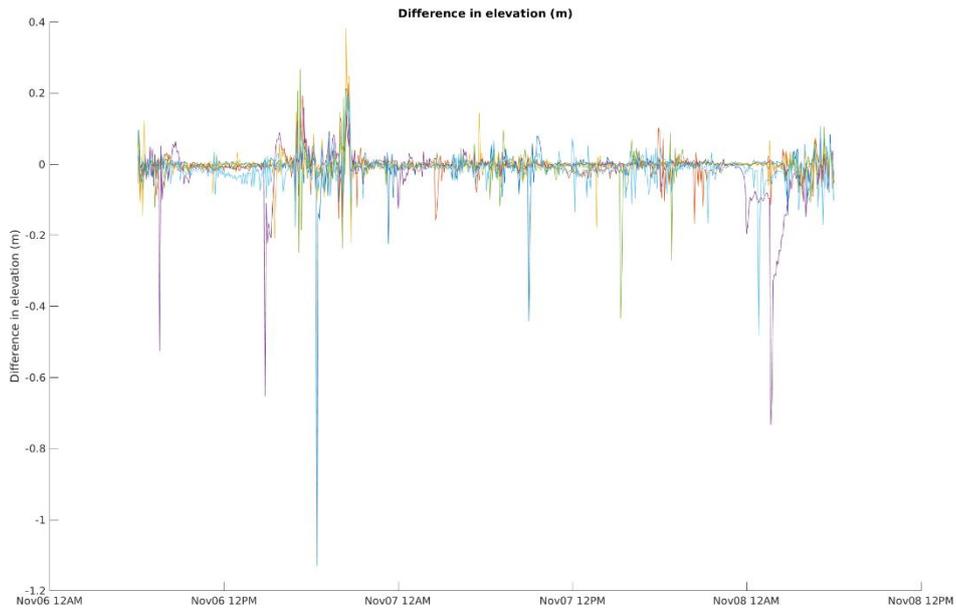

Figure 8. Stochastic simulations with additive noise $\alpha = 1.00$, $\gamma = 10^{-6}$ and its respective differences on time-series water elevation from deterministic model. Each line represents a successful simulation.

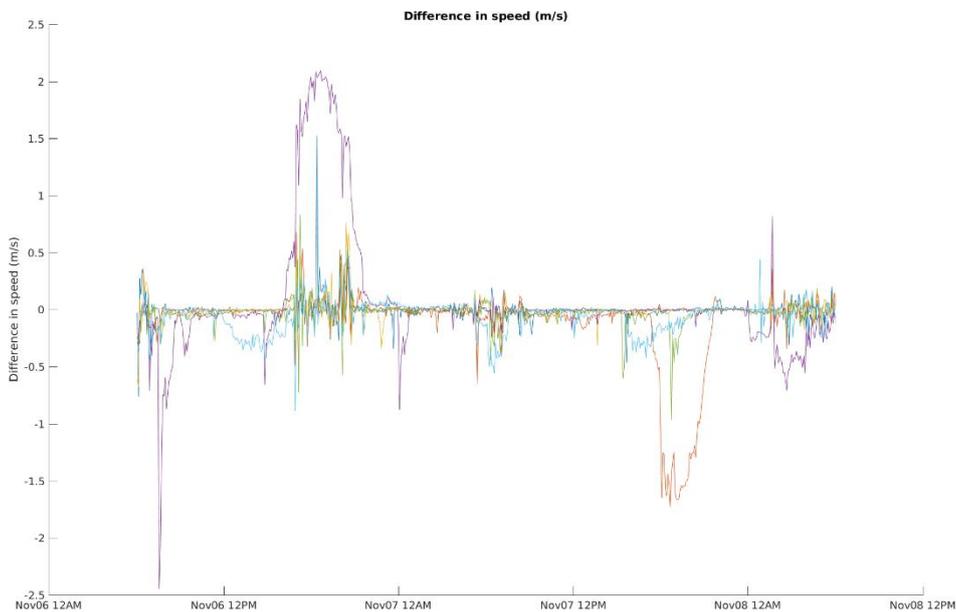

Figure 9. Stochastic simulations with additive noise $\alpha = 1.00$, $\gamma = 10^{-6}$ and its respective differences in time-series fluid speed from deterministic model. Each line represents a successful simulation.

Figures 8 and 9 shows 30 respective time-series difference from deterministic solution on water level and fluid speed when $\alpha = 1.00$, $\sigma = 10^{-6}$ (Cauchy). In this case large jumps occur more frequently so variance should be kept smaller to have successful simulations. Similar qualities from both cases were observed. Magnitude of difference in elevation becomes larger during low tide or when water is flowing relatively faster. This suggests that the model is sensitive to random external force on these two aspects. Low tide happens when the fluid speed is maximum and water elevation is lowest, driving the water away from the shore (opposite direction) without any storm surge. Higher fluid speed is brought by storm surge that causes damage to coastal areas. It is also noted that the mean of differences only averages to zero and does not accumulate. Table 1 summarizes the success rates of experiments on $\alpha$ and $\sigma$.

**Table 2. Success rates of experiments on additive stable noise for Tacloban City storm surge simulation.**

| No. of successful simulations | $\alpha = 2.00$ | $\alpha = 1.75$ | $\alpha = 1.50$ | $\alpha = 1.25$ | $\alpha = 1.00$ | $\alpha = 0.75$ | $\alpha = 0.50$ |
|---|---|---|---|---|---|---|---|
| $\sigma = 10^{-3}$ | 30/30 | | | | | | |
| $\sigma = 10^{-4}$ | 30/30 | 30/30 | 25/30 | | | | |
| $\sigma = 10^{-5}$ | 30/30 | 29/30 | 30/30 | 26/30 | | | |
| $\sigma = 10^{-6}$ | | 29/30 | 30/30 | 28/30 | 6/30 | | |
| $\sigma = 10^{-7}$ | | | | 30/30 | 26/30 | | |
| $\sigma = 10^{-8}$ | | | | | 29/30 | 0/30 | |
| $\sigma = 10^{-9}$ | | | | | 30/30 | 6/30 | 0/30 |
| $\sigma = 10^{-10}$ | | | | | 20/30 | 0/30 | |

## Conclusions

Underestimation of modeled peak heights is expected with use of bare earth topography and absence of fluid sources on the governing equations such as precipitation and river overflows. Overestimation of modeled peak heights can also occur due to presence of concrete barriers that reduced surge height and inundation extent. Using larger variance of noise $\sigma$ offers larger magnitude of difference from the deterministic model. As $\alpha$ goes to zero large jumps occur more frequently so $\sigma$ needs to be as small as $10^{-8}$ for simulation stability. The stochastic Navier-Stokes equations models storm surges better since the model is also sensitive to random external forces during low tide and relatively higher fluid speed. Since the difference of stochastic solutions from deterministic solution averages to zero it is enough to say that there is no significant improvement of the storm surge model when it comes to additive noise. This is an expected consequence since the noise used has zero mean. In the future one can consider additive and multiplicative space-correlated stable noise.

## List of abbreviations

- ADCIRC – Advanced Circulation
- SWAN – Simulating Waves Nearshore
- CFD – Computational Fluid Dynamics
- SNS – Stochastic Navier-Stokes
- CMS – Chambers, Mallows and Stuck
- DTM – Digital Terrain Model
- WGS84 – World Geodetic System 1984
- CFL – Courant-Friedrichs-Lewy
- NLCD – National Land Cover Database
- UTC – Coordinated Universal Time

[3] Brzeźniak Z, Goldys B, Le Gia Q (2015) Random dynamical systems generated by stochastic Navier-Stokes equations on a rotating sphere. Journal of Mathematical Analysis and Applications 426:505-545.

[4] Dong Z, Xu L, Zhang X (2011) Invariant measures of stochastic 2D Navier-Stokes equations driven by Lévy processes. Electronic Communications in Probability 16:678-688.

[5] Menaldi J, Sritharan S (2002) Stochastic 2-D Navier-Stokes equation. Journal of Applied Mathematics & Optimization 46:31-30.

[6] Brzeźniak Z, Hausenblas E, Zhu J (2012) 2D stochastic Navier-Stokes equations driven by jump noise. Journal of Nonlinear Analysis: Theory, Methods & Applications 79:122-139.

[7] Borak S, Hardle W, Weron R (2005) Stable distributions. Statistical Tools for Finance and Insurance 1:21-44.

[8] Chambers J, Mallows C, Stuck B (1976) A method for simulating stable random variables. Journal of the American Statistical Association 71:340-344.

[9] Siek M, Solomatine D (2010) Nonlinear chaotic model for predicting storm surges. Journal of Nonlinear Processes in Geophysics 17:405-420.

[10] Wahl T, Jensen J, Mudersbach C (2010) A multivariate statistical model for advanced storm surge analyses in the North Sea. Paper presented at the 32nd conference on coastal engineering, Shanghai, China, 30 June – 5 July 2010.

[11] Taylor A, Glahn B (2008) Probabilistic guidance for hurricane storm surge. Paper presented at the 19[th] conference on probability and statistics, New Orleans, Louisiana, 20-24 January 2008.

[12] Dietrich J, Zijlema M, Westerink J, *et. al.* (2011) Modeling hurricane waves and storm surge using integrally-coupled, scalable computations. Coastal Engineering 58:45-65.
Page **16** of **18**

## Declarations

**Availability of data and materials**

The datasets used and/or analyzed during the current study are available from the corresponding author on reasonable request.

**Competing interests**

The authors declare that they have no competing interests.

**Funding**

This research was funded by Philippines' Department of Science and Technology – Engineering Research for Development and Technology (DOST-ERDT) through research dissemination grants and full scholarship to JFR while pursuing his master's degree.

**Authors' contributions**

VPB fulfilled her role as the thesis adviser. JFR collected and preprocessed raw datasets as model inputs. JFR also modified the model's source code and use it to produce results. JFR finally wrote the manuscript with guidance from VPB. Both authors read and approved the final manuscript.

**Acknowledgements**

The authors would like to thank Dr. Amador C. Muriel for taking good time in reading this manuscript, chairing the defense panel and sharing insights about this research.